\documentclass[%
 reprint,
 amsmath,amssymb,
 aps,
]{revtex4-1}
\usepackage{qcircuit}
\usepackage{physics}
\usepackage{graphicx}% Include figure files
\usepackage{dcolumn}% Align table columns on decimal point
\usepackage{bm}% bold math
\newtheorem{theorem}{Theorem}

\usepackage{graphicx}
\usepackage[caption=false]{subfig}

\begin{document}

\preprint{APS/123-QED}

\title{Mitigating errors by quantum verification and post-selection}% Force line breaks with \\

\author{Rawad Mezher$^{1,2,\ddag}$}
\email{rawad.mezher@quandela.com}
\author{James Mills$^{1,\ddag}$}
\email{J.Mills-7@sms.ed.ac.uk}
\author{Elham Kashefi$^{1,3}$}
%\\
 \affiliation{$^1$ School of Informatics, University of Edinburgh, 10 Crichton Street, Edinburgh EH8 9AB, Scotland}
\affiliation{$^2$ Quandela SAS, 7 Rue Léonard de Vinci, 91300 Massy, France}
 %Lines break automatically or can be forced with \\\
 %\email{rawad.mezher@quandela.com; J.Mills-7@sms.ed.ac.uk}
%\author{James Mills}%
%\affiliation{1 School of Informatics, University of Edinburgh, 10 Crichton Street, Edinburgh EH8 9AB, Scotland}%Lines break automatically or can be forced with \\\
 %\email{J.Mills-7@sms.ed.ac.uk}
%\email{J.Mills-7@sms.ed.ac.uk}
%\author{Elham Kashefi}
 %\affiliation{School of Informatics, University of Edinburgh, 10 Crichton Street, Edinburgh EH8 9AB, Scotland}
 %\affiliation{$3^$ School of Informatics, University of Edinburgh, 10 Crichton Street, Edinburgh EH8 9AB, Scotland}%Lines break automatically or can be forced with \\\\\\
 %\\\
  \affiliation{$^3$ Laboratoire d’Informatique de Paris 6, Centre National de la Recherche Scientifique, Sorbonne Université, 4 place Jussieu,
75252 Paris Cedex 05, France}
\thanks{These authors contributed equally to this work.}
%\email{ekashefi@exseed.ed.ac.uk}
%\email{rawad.mezher@quandela.com; J.Mills-7@sms.ed.ac.uk}

\date{\today}% It is always \today, today,

\begin{abstract}
Correcting errors due to noise in quantum circuits run on current and near-term quantum hardware is essential for any convincing demonstration of quantum advantage. Indeed, in many cases it has been shown that noise renders quantum circuits efficiently classically simulable, thereby destroying any quantum advantage potentially offered by an ideal (noiseless) implementation of these circuits. 

Although the technique of quantum error correction (QEC) allows to correct these errors very accurately, QEC usually requires a large overhead of physical qubits which is not reachable with currently available quantum hardware. This has been the motivation behind the field of quantum error mitigation, which aims at developing techniques to correct an important part of the errors in quantum circuits, while also being compatible with current and near-term quantum hardware.

In this work, we present a technique for quantum error mitigation which is based on a technique from quantum verification, the so-called Accreditation protocol, together with post-selection. Our technique allows for correcting the expectation value of an observable $O$, which is the output of multiple runs of noisy quantum circuits, where the noise in these circuits is at the level of preparations, gates, and measurements. We discuss the sample complexity of our procedure and provide rigorous guarantees of errors being mitigated under some realistic assumptions on the noise. Our technique is tailored to time-dependent behaviours, as we allow for the output states to be different between different runs of the Accreditation protocol. We validate our findings by running our technique on currently available quantum hardware. 
\end{abstract}

\maketitle

%\tableofcontents

\section{\label{intro} Introduction}

The promise offered by quantum technologies is becoming ever closer to a reality. Recently, the so-called \emph{ quantum sampling supremacy} has been demonstrated on currently available quantum hardware \cite{arute2019quantum,zhong2020quantum}; where it was shown that particular families \cite{aaronson2011computational,boixo2018characterizing} of quantum circuits run on these hardware can outperform the best classical computers for the task of randomly sampling bit strings with a given probability distribution \cite{harrow2017quantum}. The focus now is on developing useful quantum algorithms to run on currently available and/or near-term quantum hardware (so-called noisy intermediate scale quantum (NISQ) devices \cite{preskill2018quantum}). Promising candidates in this direction are the so-called variational quantum algorithms (VQAs) \cite{cerezo2020variational}. These  consist of a classical optimizer together with a parametrized quantum circuit, and have been shown to be universal for quantum computation
\cite{biamonte2021universal}.

In practice, it is important to suppress as much as possible the noise affecting the implementation of a given quantum circuit on a given quantum hardware. Indeed, most theoretical proofs of quantum circuits outperforming their classical counterparts (the so-called \emph{quantum advantage}) require these circuits be ideal (noiseless) \cite{shor1994algorithms,grover1997quantum,bremner2011classical}, or the noise be sufficiently suppressed \cite{bremner2016average}. Moreover, it has been shown that noise in a quantum circuit can potentially destroy any quantum advantage offered by a (noiseless) version of this circuit \cite{brod2020classical,bremner2017achieving}.

A solution to the issue of noise affecting quantum circuits is the theory of quantum error correction (QEC) \cite{nielsen2002quantum,aharonov2008fault,gottesman1998theory}. In QEC a quantum circuit is composed of $n$ \emph{logical} qubits, where each such logical qubit is \emph{encoded} using $poly(n)$ \emph{physical} qubits. This allows$-$along with adequate ancilla qubit (\emph{syndrome}) measurements and classical postprocessing$-$for the error rate affecting each logical qubit to be suppressed exponentially (in $n$) \cite{nielsen2002quantum,gottesman1998theory}, provided that the error rates of individual physical qubits are below a constant value, named the fault-tolerance threshold \cite{aharonov2008fault}. Unfortunately, although QEC is fully scalable, the overhead of physical qubits required to run QEC, for circuit sizes where useful quantum advantage is expected, is still beyond the reach of current and near-term quantum hardware; though  promising  results are beginning to surface in this direction \cite{harper2019fault,nguyen2021demonstration,gertler2021protecting}.

It is crucial therefore to find techniques capable of correcting a significant amount of noise in the outputs of  quantum computations, and whose implementation can be performed on current and near-term quantum hardware. This has been the key motivation behind the field of quantum error mitigation (QEM) \cite{endo2018practical}. QEM achieves error suppression mainly by $(i)$ increasing the number of samples (runs) of a noisy quantum circuit instead of encoding qubits of this circuit as logical qubits and $(ii)$ focusing on correcting the output statistics (such as expectation values of an observable) of a quantum circuit rather than performing an active error correction of errors affecting gates, preparations and measurements of this circuit, as is done in usual QEC. These features make QEM readily implementable on available quantum hardware, as the overhead of physical qubits, quantum gates, and measurements required to implement a single circuit run with QEM is either not increased, or increased slightly as compared to the original circuit. 

There have been many proposals and techniques for performing QEM in recent years, most of which have been successfully tested experimentally on available quantum hardware. These include zero-noise extrapolation (ZNE) \cite{li2017efficient}, the quasi-probability method \cite{temme2017error}, learning based error mitigation \cite{czarnik2020error,bennewitz2021neural}, symmetry verification \cite{bonet2018low}, virtual distillation and QEM with derangement operators \cite{koczor2020exponential,huggins2020virtual}, read-out noise mitigation \cite{maciejewski2020mitigation}, and QEM by attenuating individual error sources \cite{otten2019accounting}; see \cite{endo2021hybrid} for a review of QEM techniques. Also, combinations of these techniques have been used for better error suppression \cite{endo2021hybrid}.

In this paper, we introduce a technique for QEM which is based on the \emph{Accreditation} protocol \cite{ferracin2019accrediting,ferracin2021experimentally}; a technique used to \emph{verify} the quality of the output of a desired quantum computation (see \cite{gheorghiu2019verification} for a  review of such techniques) which has recently been tested experimentally on IBM hardware \cite{ferracin2021experimentally}. In Accreditation \cite{ferracin2019accrediting,ferracin2021experimentally}, the key idea is to implement $M+1$ circuits each having a \emph{layered} structure of alternating single and two-qubit gates. One of these $M+1$ circuits is a circuit whose quality of output one would like to verify (named \emph{target} circuit), and the other $M$ circuits are so-called \emph{traps}; simplified Clifford versions of the target circuit. By running the $M$ trap circuits and examining their outputs, one can deduce the quality of the target circuit; under the assumption that the behaviour of the noise is \emph{similar} (in a sense we will specify precisely in the coming paragraphs) between traps and target. This protocol works for a broad variety of noise models \cite{ferracin2019accrediting,ferracin2019accrediting}. At the heart of the Accreditation protocol is the technique of randomized compiling \cite{wallman2016noise}, which tailors general noise affecting preparations, gates, and measurements into Pauli stochastic noise, having a simpler form and therefore being easier to analyze. 

In this work, we will use the Accreditation protocol as a technique for \emph{error detection}. A figure of merit outputted by the Accreditation protocol is an upper bound on the total variational distance ($TVD$) between the output of the target computation, and an ideal (noiseless) version of this output \cite{ferracin2019accrediting, ferracin2021experimentally}. Our idea is to set a threshold, $\varepsilon$, named \emph{quality factor}, for the $TVD$, run the Accreditation protocol multiple times each time constructing the same target circuit, and post-select on those runs where $TVD \leq \varepsilon$; that is, the runs where the figure of merit outputted by the Accreditation protocol is at most $\varepsilon$. Then, we use only these runs (i.e the outputs of the target circuits of these runs) in computing the desired expectation values. When the noise is time-dependent, what our technique allows to do is \emph{filter-out} the bad runs (where the noise levels were high) and keep only those runs where the noise levels are low enough, as defined by our quality factor. 

Under some assumptions (stated precisely in the coming paragraphs) on the time dependence of the noise, our technique provably mitigates the errors in estimating the expectation values of observables. Moreover, our technique allows for mitigating the effects of a broad range of noise models (those supported by the Accreditation protocol). Finally, our technique, unlike most techniques of QEM \cite{endo2021hybrid}, corrects errors  due to noise which is time-dependent \footnote{  Note however that the work of \cite{koczor2020exponential} also considered some time-dependent effects in QEM, see also \cite{yamamoto2021error}}.

It is important to develop error mitigation techniques  that correct time-dependent errors, especially when performing experiments on quantum hardware  over extended periods of time in order to collect enough data points. For example, in superconducting quantum hardware time-dependent errors begin to manifest after a certain period of time mainly in the form of so-called \emph{drift} errors which can significantly decrease the output quantum circuit fidelity \cite{geyko2021technical}. Furthermore, as seen in section \ref{simulation}, different subsets of qubits in a quantum device can have different noise levels. This difference in noise levels may be considered a time-dependent noise, for example when running the same quantum circuit  on different subsets of qubits of the quantum device. The technique we developed here allows for mitigating errors in both these scenarios. Also, our developed error mitigation technique could be used in conjunction with other existing error mitigation techniques which target noise that is not strongly time-dependent \cite{endo2021hybrid}; with the goal of improving overall performance. For example, in ZNE \cite{li2017efficient} a condition to be verified is that the  errors rates of the  noise channels affecting the quantum circuit should be small for the technique to give accurate results. In a scenario where multiple experiments need to be run over an extended period of time, our technique can allow detecting and discarding the instances where the above condition is not verified, then applying ZNE on the post-selected instances; thereby allowing better overall error mitigation as compared to the case where ZNE is applied alone in this scenario.

%\textbf{new par. abt imp of time dependence}

The organization of this paper is as follows. In section \ref{prelim}, we go over the Accreditation protocol, as well as the various assumptions on the noise we will adopt. In section \ref{technique} we present our technique for QEM and analyze its sample complexity. Section \ref{application} shows how we can get a provable error mitigation in scenarios of practical interest. The results of running our technique on currently available quantum hardware for the cases of a generic test circuit and a quantum circuit Born machine are in section \ref{simulation}. Finally, we discuss our results in \ref{discussion}.

\section{The Accreditation protocol and assumptions \label{prelim}}
We will adopt the Accreditation protocol used in \cite{ferracin2021experimentally}, which has been tested experimentally on IBM hardware. The difference between the work of  \cite{ferracin2021experimentally} and that of \cite{ferracin2019accrediting} is that the bounds on the $TVD$ outputted are tighter in \cite{ferracin2021experimentally}  than in \cite{ferracin2019accrediting}; however the Accreditation protocol of \cite{ferracin2019accrediting} supports more general types of noise than that of \cite{ferracin2021experimentally}.

The protocol of \cite{ferracin2021experimentally} is composed of $M+1$
circuits in total, one target circuit (whose quality of output we would like to verify), and $M$ trap circuits. Each of these circuits acts on $n$ qubits intialized in the $|0\rangle^{\otimes n}$ state. Trap and target circuits are each composed of $2m-1$ layers. Each  layer can either  be 
\begin{itemize}
    \item $type (i)$ layer: a product of Controlled $Z$  ($CZ$) gates $\prod_{\{i,j\} \in E_{l}}CZ_{i,j}$, where $E_{l}$, which we call an \emph{edge set}, is a set of qubit pairs $\{i,j\}$ on which the $CZ$ gates act, $i,j \in \{1,...,n\}$, and $CZ_{i,j}$ is a $CZ$ gate acting on qubit pair $\{i,j\}$. $E_l$ can potentially be different for different layers.
    \item $type (ii)$ layer:  a product $\otimes_{i=1,...,n} U_i$ of single-qubit gates $U_i$. If the circuit is the target circuit, then $U_i$'s are arbitrary single-qubit gates. Whereas if the circuit is a trap circuit, then $U_i$'s are all single-qubit Clifford gates, chosen according to a procedure described in \cite{ferracin2021experimentally}. These single-qubit gates both in traps and target can potentially be different for different layers.
\end{itemize}
In either trap or target circuit, there are $m$ layers of $type (ii)$, and $m-1$ layers of $type (i)$. These layers of gates are applied successively in the following alternating order: $type (ii)$ layer, followed by $type (i)$ layer, then $type (ii)$ layer, and so on. The first and last layers applied in the circuit are therefore of $type (ii)$.  All layers of $type (i)$ are exactly the same for all circuits (traps and target). The final step in the Accreditation protocol \cite{ferracin2019accrediting,ferracin2021experimentally} is measuring the qubits of all circuits in the computational ($Z$) basis. Note that a sufficiently deep  target circuit structure can implement $any$ desired quantum computation. Since the set of all single qubit gates together with one type of entangling two-qubit gate (in our case the $CZ$ gate) is enough for universal quantum computation \cite{brylinski2002universal}.

Now we will describe the idea behind how the Accreditation protocol works. In the absence of noise, the trap circuits are designed \cite{ferracin2019accrediting,ferracin2021experimentally} in a way such as to act trivially on $|0\rangle^{\otimes n}$. Thus, measuring qubits of a (noiseless) trap circuit in the $Z$ basis will always give the bit string $\{0\}:=\{0,....,0\}$. Therefore, in the presence of noise, measuring the trap circuits in the $Z$ basis and counting the number of instances where a bit string other than $\{0\}$ was obtained, one can extract information about the noise levels present in this run of the protocol, assuming these levels are \emph{similar} between all traps and target.

We will now state precisely the assumptions on the noise used in \cite{ferracin2021experimentally}, and which we will also use here.
\begin{itemize}
    \item For all circuits, preparation of qubits in the $|0\rangle^{\otimes n} $ state is followed by a preparation noise channel modelled by a completely positive trace preserving (CPTP) map $\mathcal{R}.$
    \item For all circuits, layers of $type (i)$ are followed by a CPTP map $\mathcal{E}_{CZ(E_l),l}$ representing the overall noise on this layer. $CZ(E_l)$ represents a set of $CZ$ gates with an edge set $E_l$, and $l$ is an index indicating the layer number; this noise is time (layer)-dependent, as well as gate dependent (depending on what the set $E_l$ is). 
    \item For all circuits, layers of $type (ii)$ are followed by a CPTP map of the form $\mathcal{E^{'}}_l$, where $l$ is the layer number. This noise is also time-dependent, note however that this noise is \emph{gate independent}, that is, the \emph{same} CPTP map is applied to single-qubit gates of a given layer in all circuits.
    \item For all circuits, measurements in the $Z$ basis are modelled as a CPTP map $\mathcal{M}$, representing the measurement noise, followed by an ideal measurement.
\end{itemize}

Since the layers  of $type(i)$ are the same for all circuits, then the \emph{same} CPTP maps $\{\mathcal{E}_{CZ(E_l),l}\}$ are applied to all circuits. Also, the same maps $\{\mathcal{E^{'}}_l\}$ are applied to layers of $type (ii)$ in all circuits. All circuits (traps and target) therefore experience the same types of noise, thus studying the (simpler to analyze) traps allows us directly to deduce some noise properties for the target. Finally, note that the assumption of gate independent single-qubit noise is a reasonable one in particular when looking at superconducting hardware, since the error rates of single-qubit gates are significantly less than those of two-qubit gates \footnote {see for example https://www.rigetti.com/}. Therefore, the noise on these single-qubit gates does not substantially contribute to the overall circuit noise; for sufficiently shallow circuits with small qubit numbers$-$which are the circuits  most actively being implemented experimentally at the moment.

To simplify the noise structure, randomized compiling \cite{wallman2016noise} is added on top of the Accreditation protocol. This is done by adding layers of random Paulis in between the circuit layers in traps and target according to the procedure in \cite{ferracin2021experimentally,ferracin2019accrediting,wallman2016noise}. Some additional classical postprocessing is also done in order to simplify the measurement noise \cite{wallman2016noise}. Effectively, randomized compiling \emph{twirls} the noise. That is, it transforms general CPTP maps into Pauli stochastic noise channels, with real coefficients representing the error rates. This allows to write the overall state of (trap or target) circuits at the end of the Accreditation protocol (i.e after the $Z$ basis measurements) as \cite{ferracin2021experimentally}
\begin{equation}
    \label{eqaccreditation}
    \rho_{out,i}:=(1-p_{err})\rho_{out,id,i} + p_{err} \rho_{noisy,i},
    \end{equation}
 where $i\in \{1,...,M+1\}$ is an index indicating the circuit. One of these circuits, whose index $i=\nu$ is chosen at random in the protocol \cite{ferracin2019accrediting}, is the target, all others are trap circuits. $\rho_{out,id,i}$ is the output of circuit $i$ in the absence of any noise (ideal output), $\rho_{noisy,i}$ is the state representing the effects of noise on circuit $i$. Finally, $p_{err}$ is the error rate; the probability that the noise leads to one or more errors at the level of preparations and/or gates and/or measurements. $p_{err}$ is the same for all circuits (because the same noise channels are acting on all circuits, as seen earlier). It is straightforward to see that
 \begin{equation}
     \label{eqtvd}
    TVD:=TVD(\rho_{out,\nu};\rho_{out,id,\nu}) \leq \frac{1}{2}||\rho_{out,\nu}-\rho_{out,id,\nu}||_1 \leq p_{err},
 \end{equation}
where $||.||_1$ is the usual 1-norm.

Define the probability $p_{inc}$ that a trap returns an incorrect outcome (i.e an outcome different than the $\{0\}$ string) after measurement as
\begin{equation}
    \label{eqpinc}
    p_{inc}= \lim_{M \to \infty} \frac{N_{inc}}{M},
\end{equation}
where $N_{inc}$ is the number of traps (out of $M$ total traps) which return an incorrect outcome. The authors of \cite{ferracin2021experimentally} show that $p_{err} \leq 2p_{inc} $, and therefore that
\begin{equation}
    \label{eqtvdpincrel}
    TVD \leq 2p_{inc}.
\end{equation}
By choosing a finite number 
\begin{equation}
\label{eqM}
    M \geq\frac{2ln(\frac{2}{1-\alpha})}{\theta^2},
\end{equation}

of traps,  where $\theta, \alpha \in [0,1]$, then counting the number of traps $N_{inc}$ which returned an incorrect outcome at the end of the Accreditation protocol;  and by using Hoeffding's inequality, the authors of \cite{ferracin2021experimentally} show that one can estimate $p_{inc}$ to  within $\frac{\theta}{2}$ error (in the absolute value) of its actual value (Equation (\ref{eqpinc})) with confidence greater than $\alpha$ \cite{ferracin2021experimentally}. %one can thus ensure that Equation (\ref{eqtvdpincrel}) holds with probability greater than $\alpha$, for an estimate of $p_{inc}$ within $\frac{\theta}{2}$ error (in the absolute value) of its ideal value \cite{ferracin2021experimentally}. 

Equation (\ref{eqtvdpincrel})  gives us a way of computing an upper bound on the $TVD$ by using the outputs of trap circuits in the accreditation protocol, for a suitable number of such traps given by Equation (\ref{eqM}).

So far, we have discussed a single run of the Accreditation protocol, which involves preparing and measuring $M+1$ trap and target circuits. Our goal, as stated in section \ref{intro}, is to run the Accreditation protocol multiple times (with the same target circuit for each run), and keep only the runs where $TVD \leq \varepsilon$, for some $\varepsilon \in [0,1]$ we define, and which we call quality factor. In between different runs of the Accreditation protocol, we will allow our noise to \emph{vary}. That is, we allow $p_{err}$ and the states $\{\rho_{noisy,i}\}_{i=1,..,M+1}$ (see Equation (\ref{eqaccreditation})) to vary between runs. Thereby allowing us to treat time-dependent behaviours. We will denote ${p^{j}}_{err}$ and $\{ {\rho^{j}}_{noisy,i}\}_{i=1,..M+1}$ to indicate the value of $p_{err}$, and the states $\{\rho_{noisy,i}\}_{i=1,..M+1}$ at run $j$ (see Equation (\ref{eqaccreditation})). We will now make the following two assumptions on our time-dependent noise behaviour.
\begin{itemize}
    
    \item  $A1$: The number of possible noise behaviours is \emph{finite}. That is, for any run $j$ of the Accreditation protocol,  the couple $\Big\{{p^{j}}_{err},\{ {\rho^{j}}_{noisy,i}\}_{i=1,...,M+1}\Big\}$ describing the noise behaviour at run $j$ can only be one of a set $N$ of distinct such couples
    \begin{multline*}
        \Big\{ \{p_{err,1},\{\rho_{noisy,i,1}\}_{i=1,..,M+1}\} ;...\\;\{p_{err,N}, \{\rho_{noisy,i,N}\}_{i=1,...,M+1}\} \Big\},
    \end{multline*}
    where $N$ is a finite positive integer, and $\{p_{err,l},\{\rho_{noisy,i,l}\}\}$ indicates a possible noise behaviour indexed $l$, where $l \in \{1,..,N\}$. 
    \item $A2$: The noise behaviours are distributed \emph{uniformly} (each behaviour appearing with probability $\frac{1}{N}$) and \emph{independently} between runs.
\end{itemize}

We close this section with a few remarks on our assumptions on the time-dependent behaviour of the noise. The first is that $A1$ is required in general for the expectation values to converge to a fixed value (see section \ref{technique} for more details).
%;  however for noise which is very close to depolarizing, and for mitigating expectation values of Pauli observables this assumption is not required (see section \ref{application}). Note that a behaviour of noise which is almost depolarizing has been observed in many experiments performed on currently available quantum hardware \cite{ville2021leveraging,urbanek2021mitigating}. 
The requirement of uniform distribution of the noise in $A2$ is not required in general, however we use it to simplify the analysis in section \ref{technique}.  The requirement of independence in $A2$ is needed to simplify deriving Chernoff-type bounds for the convergence rates, as well as estimate the sample complexity (see section \ref{technique} and appendix \ref{appendix} for more details).
%\bigskip
\section{ Error mitigation protocol \label{technique}}

The overall goal of our work is to compute an, as close as possible, approximation of the expectation value of some observable $O$ with respect to the ideal (noiseless) state of the target circuit. Although our technique works for general observables, in the rest of this paper we will consider, for simplicity of analysis, only Pauli observables of the form
\begin{equation}
    \label{eqobservable}
    O=\bf{p}_1 \otimes \bf{p}_2 ...\otimes \bf{p}_n,
\end{equation}
where $\bf{p_i} \in \{1,X,Y,Z\}$ for $i=1,..,n$, and  $O \neq \bf{1}^{\otimes n}$; $\bf{1}$ is the single-qubit identity, and $\bf{X}$, $\bf{Y}$ and $\bf{Z}$ 
are the usual single-qubit Pauli matrices. Note that any such observable can be measured by incorporating for example a layer of single-qubit gates, call it $U^{\dagger}_{O}$, into the final layer of $type(ii)$ before the $Z$ basis measurements in the target circuit, then postprocessing after the measurement to transform the measurement outcome onto the desired basis (i.e an eigenstate of $O$). Finally,  note that our choice to work with only Pauli observables is not necessarily a limitation; since, by linearity of the expectation value, we can also compute the expectation value of a general observable, which is a linear combination of Pauli observables $\sum_{i=1,..,k}\alpha_iO_i$, where $\alpha_i$ are real numbers and $O_i$ are Pauli observables. We can do this  by computing the expectation value of each $O_i$ individually as is done for example in \cite{paini2021estimating}. In particular, if $k$ scales efficiently with the system size and the expansion coefficients are easily determinable, then the process of estimating $<\sum_{i=1,..,k}\alpha_iO_i>$ does not introduce substantial overhead as compared to estimating a single Pauli observable $<O>$.

We will rewrite the ideal target circuit output (see Equation (\ref{eqaccreditation}))  as
\begin{equation}
    \label{eqidealtarget}
    \rho_{out,id,\nu}=\sum_{s}p_{s,id}|s\rangle \langle s|,
\end{equation}
where $p_{s,id}$ is the probability that the $Z$ basis measurement gives rise to a bit string $s \in \{0,1\}^{n}$, and the sum ranges over all $2^n$ possible bit strings $s$.
Also,
\begin{equation}
    \label{eqnoisytarget}
    \rho_{noisy,\nu,l}=\sum_{s}p_{n,s,\nu,l}|s\rangle \langle s|,
\end{equation}
where $p_{n,s,\nu,l}$ is the probability of obtaining outcome $s \in \{0,1\}^n$ after the $Z$ measurements, when the target circuit experiences noise, for a noise behaviour $l \in \{1,...,N\}$. The sum ranges over all $2^n$ bit strings $s$. 

Since $O$ is a Pauli operator, it can have two possible eigenvalues, +1 and -1, and $Tr(O)=0$. Let $\mathcal{S}_1$ be the set of states $\{|s\rangle\}$ with $s \in \{0,1\}^n$ such that $OU_{O}|s\rangle=U_{O}|s\rangle$; here $U_{O}$ is the inverse of $U^{\dagger}_{O}$, which was added in the target circuit just before the $Z$ basis measurements in order to measure $O$ (see beginning of this section). Note that $U_O$ transforms the computational basis state $|s\rangle$ onto an eigenstate of $O$. Similarly, let $\mathcal{S}_{-1}$ be the set of states $\{|s\rangle\}$ where $s \in \{0,1\}^n$ such that $OU_{O}|s\rangle=-U_{O}|s\rangle$. From the properties of Pauli operators, $|\mathcal{S}_1|=|\mathcal{S}_{-1}|=2^{n-1}$. The expectation value $<O>_{id}$ of $O$ for an ideal (noiseless) target circuit is then
\begin{equation}
    \label{eqidealexpectation}
    <O>_{id}=\sum_{s \in \mathcal{S}_1}p_{s,id}-\sum_{s' \in \mathcal{S}_{-1}}p_{s',id}.
\end{equation}
The expectation value $<O>_l$ of $O$ in the state of a target circuit with noise behaviour $l \in \{1,...,N\}$ is 
\begin{multline}
    \label{eqexpectationnoisebehaviourj}
    <O>_l=\sum_{s \in \mathcal{S}_1}(1-p_{err,l})p_{s,id}+p_{err,l}p_{n,s,\nu,l}\\ - \Big (\sum_{s' \in \mathcal{S}_{-1}} (1-p_{err,l})p_{s',id}+p_{err,l}p_{n,s',\nu,l} \Big).
\end{multline}
The expectation over all possible noise behaviours is given by (for a uniform distribution over noise behaviours, see assumption $A2$ in section \ref{prelim})
\begin{equation}
    \label{eqnoisyexpectation}
    <O>_{noisy}=\frac{1}{N}\sum_{l=1,...N}<O>_{l}.
\end{equation}
Suppose there are $\omega < N$ noise behaviours satisfying
\begin{equation}
    \label{eqTVD2}
    TVD_l:=TVD(\rho_{out,\nu,l};\rho_{out,id,\nu}) \leq \varepsilon,
\end{equation}
where $l \in \{1,...,\omega\}$ indicates the noise behaviour and
\begin{equation}
    \label{eqrhooutnoisebehaviourl}
    \rho_{out,\nu,l}=(1-p_{err,l})\rho_{out,id,\nu}+p_{err,l}\rho_{noisy,\nu,l},
\end{equation}
is the output state of the target for the noise behaviour $l$, and $\varepsilon$ is our pre-defined quality factor. 

Our error mitigation protocol estimates the following expectation value
\begin{equation}
    \label{eqerrormitigatedexpectation}
    <O>_{mit}=\sum_{l=1,..,\omega}\frac{1}{\omega}<O>_l,
\end{equation}
where the sum is over all noise behaviours with $TVD_l \leq \varepsilon$  (see Equation (\ref{eqTVD2})). Note that, because of our assumption $A1$ (see section \ref{prelim}), both $<O>_{mit}$ and $<O>_{noisy}$ converge to a fixed value in the interval $[-1,1]$.

Our error mitigation protocol can be described as follows:
%\bigskip
\begin{itemize}
    \item Run $K$ times the Accreditation protocol with $M$ trap circuits (these circuits can potentially be different between different runs) and the same target circuit, where $M$ is given by Equation (\ref{eqM}). For each run $j=1,...,K$, let $\lambda_j \in \{-1,+1\}$ be the eigenvalue obtained after measuring $O$ on the target circuit at this run and getting an output $|s\rangle \in \mathcal{S}_1 \cup \mathcal{S}_{-1}$.
    \item For each run $j=1,...,K$, compute an upper bound for $TVD$  using the procedure described in section \ref{prelim} (see Equation (\ref{eqtvdpincrel})), and keep only those runs where $TVD \leq \varepsilon$ \footnote{ That is, when our estimated upper bound on $TVD$ is $\leq \varepsilon$}.
    \item If $m \leq K$ is the number of runs where $TVD \leq \varepsilon$, compute
    \begin{equation}
    \label{eqestimateomit}
    <\tilde{O}>_{mit}=\sum_{j=1,...,m}\frac{1}{m}\lambda_{j},
    \end{equation}
    where the sum ranges over all runs with $TVD \leq \varepsilon$. For sufficiently large $K$, we obtain a good estimate of $<O>_{mit}$, that is, for large enough $K$, the approximation
    \begin{equation}
        \label{eqestimateomit2}
        <\tilde{O}>_{mit} \approx <O>_{mit},
    \end{equation}
    is accurate.
    
\end{itemize}

Having presented our error mitigation protocol, we will now study its sample complexity as well as analyze its convergence properties. For a noise behaviour $l=1,...,N$ define
\begin{multline}
    \label{eqpldef}
    p_{l}(+1):=\sum_{s \in \mathcal{S}_1}(1-p_{err,l})p_{s,id}+p_{err,l}p_{n,s,\nu,l}\\
    p_{l}(-1):=\sum_{s' \in \mathcal{S}_{-1}}(1-p_{err,l})p_{s',id}+p_{err,l}p_{n,s',\nu,l}.
\end{multline}
in this case, $<O>_{l}$ in Equation (\ref{eqexpectationnoisebehaviourj}) can be rewritten as
$$<O>_{l}=p_{l}(+1)-p_{l}(-1).$$
For a run $j=1,...,K$, let $\pi(j) \in \{1,...,N\}$ be an index labelling the noise behaviour that appeared at this run, and let  $p^{j}_{\pi(j)}(+1)$ (similarly $p^{j}_{\pi(j)}(-1)$) denote the quantity $p_{\pi(j)}(+1)$ (similarly $p_{\pi(j)}(-1)$ )  at this run (Equation (\ref{eqpldef})).

Define
\begin{equation}
    \label{eqsw}
    s_{\omega}:=\frac{1}{m}\sum_{j=1,...m}p^{j}_{\pi(j)}(+1)-p^{j}_{\pi(j)}(-1),
\end{equation}
where the sum ranges over all runs $m \leq K$ where we obtained a noise behaviour with $TVD \leq \varepsilon$, and $\pi(j) \in \{1,...,\omega\}$. Furthermore, let 
\begin{equation}
    \label{eqvariance}
    \sigma_{\omega}=\sqrt{\sum_{l=1,...,\omega}\frac{1}{\omega}(p_{l}(+1)-p_{l}(-1)-<O>_{mit})^2},
\end{equation}
where the sum ranges over all noise behaviours with $TVD_{l} \leq \varepsilon$ (Equation (\ref{eqTVD2})). Finally, let $\varepsilon_{1}, \gamma \in [0,1]$. We will mean by $Pr(A)$ the probability that event $A$ occurs. We first show that
\begin{theorem}
\label{th1}
$Pr(|s_{\omega}-<O>_{mit}| \leq \varepsilon_{1}) \geq \gamma$, when 
\begin{equation}
\label{eqK}
K \approx \geq \frac{N.\sigma^{2}_{\omega}}{\omega(1-\gamma)\varepsilon^{2}_1}, 
\end{equation}
where (\ref{eqK}) holds approximately in the limit of a large number of runs.
\end{theorem}
\bigskip
 We constrain $ 2\lambda-1\leq s_{\omega} <1$, where $\lambda$ can be an arbitrarily small (but non-zero) positive constant. Also, define $\beta:=\beta(m)$ a function of $m$ such that
$\frac{1}{2}\leq \beta<1$ when $s_\omega \geq 0$; and $1<\beta \leq \frac{1}{2\lambda}$ when $s_{\omega}<0$. Our second result is the following theorem
\begin{theorem}
\label{th2}
For $s_{\omega} \geq 0$ the following inequalities hold
\begin{equation}
\label{eqth21}
Pr(<\tilde{O}>_{mit} \leq (1+\delta)s_{\omega}) \geq 1-e^{-l},
\end{equation}
\begin{equation}
\label{eqth22}
Pr(<\tilde{O}>_{mit} \geq (1-\delta)s_{\omega}) \geq 1-e^{-l},
\end{equation}
where $l>>1$ is a positive integer and
$$\delta \approx \frac{\sqrt{2\beta} }{1-\beta}\sqrt{\frac{l\varepsilon^{2}_1(1-\gamma)}{\sigma^2_{\omega}}}. $$
Similarly, for $s_{\omega}<0$ the following inequalities hold
\begin{equation}
\label{eqth23}
Pr(<\tilde{O}>_{mit} \geq (1-\delta')s_{\omega}) \geq 1-e^{-l'},
\end{equation}
\begin{equation}
\label{eqth24}
Pr(<\tilde{O}>_{mit} \leq (1+\delta')s_{\omega}) \geq 1-e^{-l'},
\end{equation}
with $$\delta' \approx \frac{\sqrt{2\beta}}{\beta-1}\sqrt{\frac{l'\varepsilon^{2}_1(1-\gamma)}{\sigma^2_{\omega}}},$$ subject to the constraint that $$0<l' \leq (\frac{2\lambda}{1-2\lambda})^2\frac{\sigma^2_{\omega}}{(1-\gamma)\varepsilon^{2}_1}.$$
\end{theorem}
We prove both Theorem \ref{th1} and \ref{th2} in the appendix \ref{appendix}.
%\bigskip

Together, Theorem \ref{th1} and \ref{th2} guarantee that the estimate $<\tilde{O}>_{mit}$ (Equation (\ref{eqestimateomit})) of the error mitigated expectation value converges to  $<O>_{mit}$ (Equation (\ref{eqerrormitigatedexpectation})) for large enough number of runs $K$. We show this  by first showing the convergence of $<\tilde{O}>_{mit}$ to an intermediate quantity $s_{\omega}$ (Theorem \ref{th2}), then showing the convergence of $s_{\omega}$ to $<O>_{mit}$ (Theorem \ref{th1}). It is worth noting that the convergence rate and sample complexity of our procedure depend, in addition to their dependence on the precision ($\varepsilon_1$) and confidence ($\gamma$), on the \emph{type} of noise behaviour, as quantified by the probabilities $\{p_{n,s,\nu,l}\}$ and $\{p_{err,l}\}$. This can be seen  through the dependence of $K, \delta, \delta'$ on the quantity $\sigma^2_{\omega}$ which can be thought of as a type of variance.

With Equations (\ref{eqK}) and (\ref{eqM}) in hand, we can now compute the sample complexity of our procedure. The total number of circuits $C_{tot}$ (traps and targets) over all runs needed to implement our error mitigation protocol is 
\begin{equation}
    \label{eqcircuitstotal}
    C_{tot}=K(M+1) \geq \frac{N.\sigma^{2}_{\omega}}{\omega(1-\gamma)\varepsilon^{2}_1}\big(\frac{2ln(\frac{2}{1-\alpha})}{\theta^2} +1 \big).
\end{equation}
As seen previously, $M+1$ is the number of circuits needed for a single run of the Accreditation protocol, and $K$ is the number of runs of the Accreditation protocol required to estimate $<O>_{mit}$ to a good precision.

\begin{figure}[t]
\hspace*{-2cm}  
\includegraphics[scale=0.55]{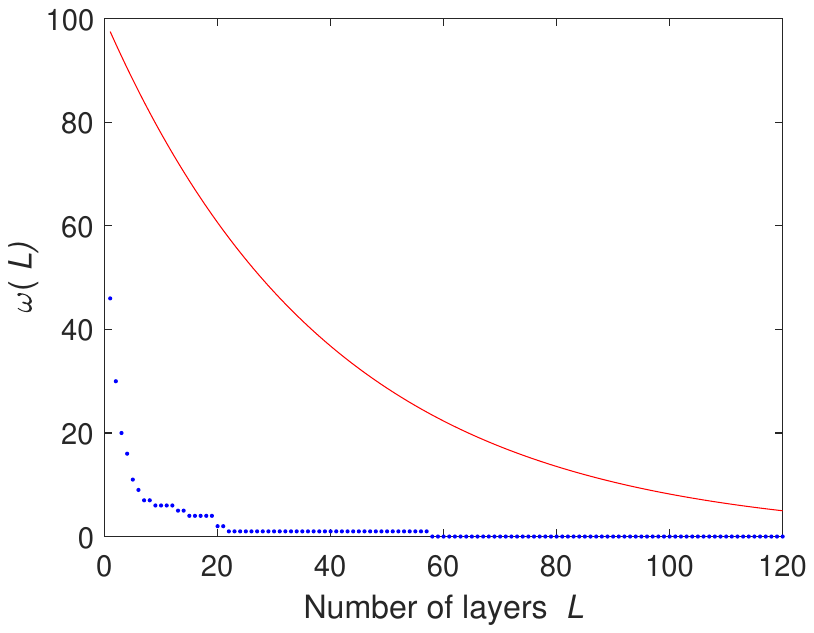}
\centering
\vspace*{-1.6cm} 
\caption{$\omega:=\omega(L)$ in function of the number of layers $L$ of the target circuit. We study the case of $N$ depolarizing noise behaviours, where the depolarizing strength per layer is $\epsilon_i$ for noise behaviour $i \in \{1,...,N\}$, and is the same for all layers. In this figure, $N=100$ and $\varepsilon=0.5$. The lower blue (dotted) curve represents the values of $w(L)$ for each $L$, the upper red (solid) curve is a decreasing exponential $f(L)=Ne^{-0.025L}$ upper bounding $\omega(L)$. In our numerics, we did not specify the number of qubits $n$ of the circuit. In practice, $n$ affects the values of $\epsilon_i$ as well as the value of $||\rho_{id}-\frac{\mathbf{1}_n}{2^n}||_1$, where $\rho_{id}$ is the ideal state of the target circuit. In our calculations we choose $||\rho_{id}-\frac{\mathbf{1}_n}{2^n}||_1 \approx 2$, and $\epsilon_i$ to be distributed uniformly randomly in $[0,1]$, as seen in the main text.}
\label{fignumerics}
\end{figure}

It is worth discussing our sample complexity in light of the result of \cite{takagi2021fundamental}. In \cite{takagi2021fundamental}, it was shown that, in the case of depolarizing noise, the lower bound on the sample complexity of a broad family of error mitigation protocols increases exponentially with the number of layers $L$ of the noisy quantum circuit being mitigated. The expression for the sample complexity $C_{tot}$ of our procedure (Equation (\ref{eqcircuitstotal})) points to a similar conclusion, although the exponential dependence on $L$ is not explicitly present in (\ref{eqcircuitstotal}). To see this, note that for a fixed $\varepsilon$, the number of noise behaviours $\omega$ with $TVD \leq \varepsilon$ decreases with increasing number of layers $L$. Indeed, for a depolarizing noise channel acting on every layer of our target circuit,  the overall noise (whose increase causes an increase in $TVD$) increases with increasing $L$, therefore the number of noise behaviours $\omega:=\omega(L)$ decreases with increasing $L$; leading to an increase in $C_{tot}$ which is inversely proportional to $\omega$ (Equation (\ref{eqcircuitstotal})). To give evidence of the exponential (in $L$) increasing behaviour of $C_{tot}$, we performed numerical calculations   for $N$ depolarizing noise behaviours with noise strengths per layer $\epsilon_i$, $i \in \{1,...,N\}$ (for each noise behaviour $i$, the noise strength is the same for all layers and is $\epsilon_i$, the noise channel acting per layer of the circuit has the form $\Lambda_i(\rho)=(1-\epsilon_i)\rho+\epsilon_i \frac{\mathbf{1}_n}{2^n}$, $\mathbf{1}_n$ is the identity on $n$ qubits.); where $\epsilon_i$ are uniformly randomly distributed in the interval $[0,1]$. Thus, after $L$ layers the noise behaviour $i \in \{1,...,N\}$ induces a depolarizing channel with noise strength $1-(1-\epsilon_i)^L$, as can be seen directly by applying $L$ times the depolarizing channel $\Lambda_i$. For a fixed $\varepsilon$, these calculations show that $\omega(L)$ is a decreasing (with intervals of being constant) function  upper bounded by an exponentially decreasing function of $L$ (therefore $C_{tot}$ is lower bounded by an exponentially increasing function with $L$, similar to the results in \cite{takagi2021fundamental}) up until $\omega(L)=0$ is reached; see Figure \ref{fignumerics}. After $\omega(L)=0$, our technique no longer works, as there are no more behaviours to post-select over. Finally, note that our reasoning is valid  in the case when $\omega:=\omega(L)>1$ (if $\omega=1$, then $\sigma_{\omega}=0$ and the lower bound on $C_{tot}$ in (\ref{eqcircuitstotal}) is trivial), and for noise behaviours satisfying $\sigma_{\omega}:=\sigma_{\omega}(L) > \kappa \neq 0$, where $\kappa=min_{L} (\sigma_{\omega}(L)) \in [0,2]$ is a constant. Our numerical results suggest that, similar to other error mitigation protocols \cite{endo2021hybrid}, our technique will likely give good results, and have a reasonable sample complexity, for low-to-moderate depth quantum circuits.
 
Intuitively, the error mitigated expectation value \\ $<O>_{mit}$ (Equation (\ref{eqerrormitigatedexpectation})) should be closer to the noiseless expectation value $<O>_{id}$ (Equation (\ref{eqidealexpectation})) than the unmitigated expectation value $<O>_{noisy}$ (Equation (\ref{eqnoisyexpectation})). This is because $<O>_{noisy}$ ranges over \emph{all} noise behaviours, including those behaviours where the noise levels are high ($TVD > \varepsilon$). Whereas, $<O>_{mit}$ filters out these high noise level behaviours by only considering behaviours where $TVD \leq \varepsilon$. Therefore, we expect the following relation (which shows our procedure mitigates errors) to hold in general
\begin{equation}
    \label{eqmitigationinequality}
    |<O>_{mit}-<O>_{id}| \leq |<O>_{noisy}-<O>_{id}|.
\end{equation}
However, proving Equation (\ref{eqmitigationinequality}) is difficult without any a priori knowledge of the noise. In the next section, we will prove Equation (\ref{eqmitigationinequality}) for the case of a depolarizing noise behaviour which has been observed frequently in experiments run on quantum hardware \cite{ville2021leveraging,urbanek2021mitigating}.
\section{\label{application} Provable error mitigation for the case of depolarizing noise}
In this section, we will prove Equation (\ref{eqmitigationinequality}) for the case of depolarizing noise. As mentioned previously, a noise behaviour which is dominantly depolarizing has been observed in multiple recent experiments \cite{ville2021leveraging,urbanek2021mitigating}, and therefore showing that our technique provides a provable error mitigation in this case is important.

For depolarizing noise, the followng holds for all $l=1,...,N$,
$\rho_{noisy,\nu,l}= \frac{\mathbf{1}_n}{2^{n}}$, where $\mathbf{1}_n$ is the identity on $n$-qubits. Therefore,
$p_{n,s,\nu,l}=\frac{1}{2^n},$ for all $s \in \{0,1\}^n$ and $l=1,...,N$. Plugging this into Equation (\ref{eqexpectationnoisebehaviourj}) while noting that for Pauli observables $O$, $|\mathcal{S}_{1}|=|\mathcal{S}_{-1}|=2^{n-1}$, we obtain straightforwardly that
$$<O>_{l}=(1-p_{err,l})<O>_{id}.$$
for all $l=1,..,N$. Using this we get that  $<O>_{noisy}$ and $<O>_{mit}$ for depolarizing noise become (see Equations (\ref{eqnoisyexpectation}) and (\ref{eqerrormitigatedexpectation}))
$$<O>_{noisy}=(1-\sum_{l=1,..,N}\frac{p_{err,l}}{N})<O>_{id},$$ and
$$<O>_{mit}=(1-\sum_{l=1,..,\omega}\frac{p_{err,l}}{\omega})<O>_{id}.$$
Therefore
$$|<O>_{noisy}-<O>_{id}|=\sum_{l=1,..,N}\frac{p_{err,l}}{N}|<O>_{id}|,$$ and
$$|<O>_{mit}-<O>_{id}|=\sum_{l=1,..,\omega}\frac{p_{err,l}}{\omega}|<O>_{id}|.$$
Now $p_{err,l} \leq 2p_{inc,l} \leq \varepsilon$ (see section \ref{prelim}) for $l=1...,\omega$, therefore
$$\sum_{l=1,..,\omega}\frac{p_{err,l}}{\omega} \leq \varepsilon.$$

We can rewrite $$\sum_{l=1,..,N}\frac{p_{err,l}}{N}=\frac{\omega}{N}\sum_{l=1,..,\omega}\frac{p_{err,l}}{\omega}+\frac{N-\omega}{N}\sum_{l=\omega+1,..N}\frac{p_{err,l}}{N-\omega}.$$ We have used the convention that  $l=1,...,\omega$ are the noise behaviours with $TVD_{l} \leq \varepsilon$ and $l=\omega+1,..,N$ are those where $TVD_{l}>\varepsilon$.
Thus, $$\sum_{l=\omega+1,..N}\frac{p_{err,l}}{N-\omega} > \varepsilon \geq \sum_{l=1,..\omega}\frac{p_{err,l}}{\omega}.$$ Plugging this into the above expression we obtain
\begin{multline*}
    \sum_{l=1,..,N}\frac{p_{err,l}}{N}>\frac{\omega}{N}\sum_{l=1,...\omega}\frac{p_{err,l}}{\omega}+\frac{N-\omega}{N}\sum_{l=1,...\omega}\frac{p_{err,l}}{\omega} \\ > \sum_{l=1,..,\omega}\frac{p_{err,l}}{\omega} .
\end{multline*}

Therefore 
$$\sum_{l=1,..,\omega}\frac{p_{err,l}}{\omega}|<O>_{id}| <\sum_{l=1,..,N}\frac{p_{err,l}}{N}|<O>_{id}|, $$
and Equation (\ref{eqmitigationinequality}) is proven for the case of depolarizing noise.

\section{\label{simulation}Running our error mitigation protocol on quantum hardware}

Time-dependent effects in noise are often neglected when assessing quantum device performance, however it is an important consideration to take into account when running algorithms that require sampling from a device over an extended time period. Currently available superconducting devices are periodically tuned while they are online and potentially running quantum software applications. This prevents significant device performance deterioration with time but also results in a source of continual fluctuation in device noise. One type of event this tuning protects against is readout classifier drift, whereby the readout fidelity degrades with time and gradually introduces bias and therefore error to the quantum measurement \cite{geyko2021technical}. Another is coherent gate error \cite{koczor2020exponential} due to imperfectly calibrated gate operations; however note that such errors can also be accounted for and mitigated in our technique because of the randomized compiling \cite{wallman2016noise} which transforms these errors into stochastic Pauli errors. 

Another relevant scenario, which is the focus of the next sections, is if one is running experiments on different subsets of qubits on a single device, in this case one may consider the difference in noise a time-dependent effect in the context of time-dependent noise mitigation. These time-dependent effects mentioned and many more combine to give a complex and ever-changing source of errors which are unique to each device and also to each experiment separated in time.

\subsection*{Hardware Description}

We ran our experiments on Rigetti Computing's quantum hardware.  We used two quantum processing units (QPUs), Rigetti's Aspen-9 and Aspen-11. These devices are composed of transmon qubits with alternating  octagonal and square topologies \cite{hong2020demonstration,abrams2020implementation,reagor2018demonstration} (see Figures \ref{fig4} (a) and (b)). Aspen-9 is composed of 32 qubits with a median thermal
relaxation time (T1) of 33$\mu$s, median dephasing time (T2) of 16$\mu$s \cite{youssef2020measuring}, a median single-qubit gate fidelity of 99.39\%, and a median two-qubit gate fidelity of 94.28\%. The single-qubit native gates for this device are $RZ(\theta)$ and $RX(\pi/2)$, where the first is a local rotation by $\theta$ around the Z axis and the second a rotation by $\pi/2$ about the X axis. The two-qubit native gates are $CZ$ and $XY$, which are respectively a controlled $Z$ gate and a parametrized iSWAP gate \footnote{https://pyquil-docs.rigetti.com/en/v2.28.0}. Aspen-11 is composed of 38 qubits with a median T1 time of 29$\mu$s, T2 time of 16$\mu$s, a median single-qubit gate fidelity of 99.81\% and a median two-qubit gate fidelity of 94.27\%. The native gates of this device are the same as those of Aspen-9. We have programmed our error mitigation technique to  run  on these devices by using the Rigetti stack, and the associated tool-kits of the $\mathsf{Quil}$ language as well as the optimizing compiler $\mathsf{Quil-C}$ \cite{karalekas2020quantum, smith2016practical,smith2020open}. 

Further details regarding the specifications of the qubits used for our experiments on Aspen-9 and Aspen-11 are included in Appendix \ref{appB}
%Details regarding the device
%specifications  for Aspen-9 and Aspen-11 are included in Appendix B.

\subsection*{Generic Test Circuit Experiment}

\begin{figure}[h]
\includegraphics[scale=0.28]{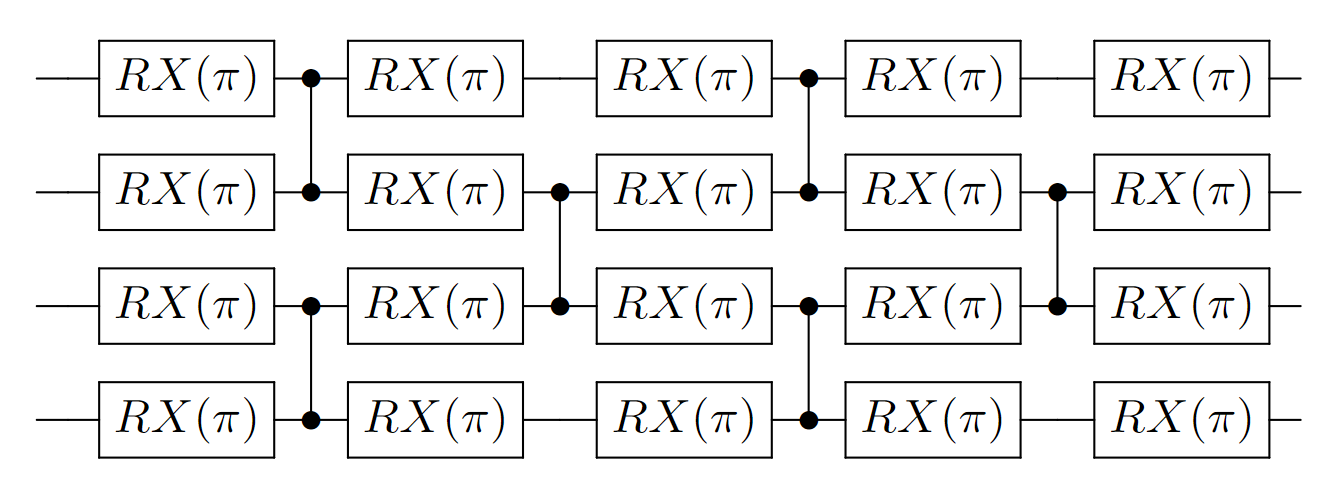}
\centering
% \begin{align*}
% \scalebox{0.9}{
% \Qcircuit @C=0.65em @R=0.8em {
%  &\qw &\gate{RX(\pi)} & \ctrl{1} & \gate{RX(\pi)} & \qw &\gate{RX(\pi)} & \ctrl{1} & \gate{RX(\pi)}  & \qw &\gate{RX(\pi)} & \qw\\
%  &\qw & \gate{RX(\pi)} & \control \qw & \gate{RX(\pi)} & \ctrl{1} & \gate{RX(\pi)}  & \control \qw & \gate{RX(\pi)} & \ctrl{1} & \gate{RX(\pi)} & \qw\\
%  &\qw & \gate{RX(\pi)} & \ctrl{1}  & \gate{RX(\pi)} & \control \qw & \gate{RX(\pi)} & \ctrl{1} & \gate{RX(\pi)} & \control \qw & \gate{RX(\pi)} & \qw\\
%  &\qw &\gate{RX(\pi)} & \control \qw & \gate{RX(\pi)}& \qw &\gate{RX(\pi)} & \control \qw & \gate{RX(\pi)} & \qw &\gate{RX(\pi)} & \qw\\
% }
% }
% \end{align*}
\caption{An example of the target circuit used in the experiment with four qubits and nine gate layers. When run as part of the Accreditation protocol the single-qubit gates are compiled together with the single-qubit Pauli gates included in the quantum one-time pad, however this does not change the overall computation.}
\label{fig1}
\end{figure}

To demonstrate the mitigating effect of our technique against time-dependent errors (here corresponding to difference in noise levels between two subsets of qubits of a quantum device), we first ran a generic experiment on the Aspen-9 QPU.  %we ran it on the 32 qubit Aspen-9 Rigetti superconducting quantum processing unit (QPU). 
%The device has an octagonal topology with a median T1 time of 33$\mu$s, median T2 time of 16$mu$s, a median single-qubit gate fidelity of 99.39\% and a median 2 qubit gate fidelity of 94.28\%. The single-qubit native gates for this device are $RZ(\theta)$ and $RX(\pi/2)$, where the first is a local rotation by $\theta$ around the Z axis and the second a rotation by $\pi/2$ about the X axis. The two-qubit native gates are $CZ$ and $XY$, that is a controlled-Z operation and a rotation operation between the $|01\rangle$ and $|10\rangle$ states. 
During the experiment we used  circuits with alternating layers of $RX(\pi)$ gates (which are implemented on the device as two $RX(\pi/2)$ gates) and $CZ$ gates. These circuits act on four qubits. An example  with nine layers of alternating  $RX(\pi/2)$ and $CZ$ gate layers is shown in Figure \ref{fig1}. The single-qubit $RX(\pi)$ gates within each single-qubit gate layer are compiled together with the single-qubit quantum one-time pad gate operations (random single-qubit Pauli gates which are added in order to implement randomized compiling \cite{wallman2016noise,ferracin2019accrediting}), this randomizes these single-qubit operations  while keeping the overall computation unchanged \cite{wallman2016noise}. This general circuit structure of alternating layers of single-qubit rotation and two-qubit gates is widely used in NISQ-appropriate algorithms in for example both quantum machine learning and quantum chemistry simulation \cite{wei2020full,biamonte2017quantum}. Circuits with this overall structure can be computationally universal depending on the choice of the single-qubit rotation gates. Our chosen example circuit is similar to the commonly used quantum machine learning (QML) \cite{biamonte2017quantum} anstatz circuit  , where instead of fixing the angle of rotation one would have an array of rotation angles which would be iteratively updated during training to optimize the circuit with respect to 
a given task.

\begin{figure}[t]
\hspace*{0cm}  
\includegraphics[scale=0.37]{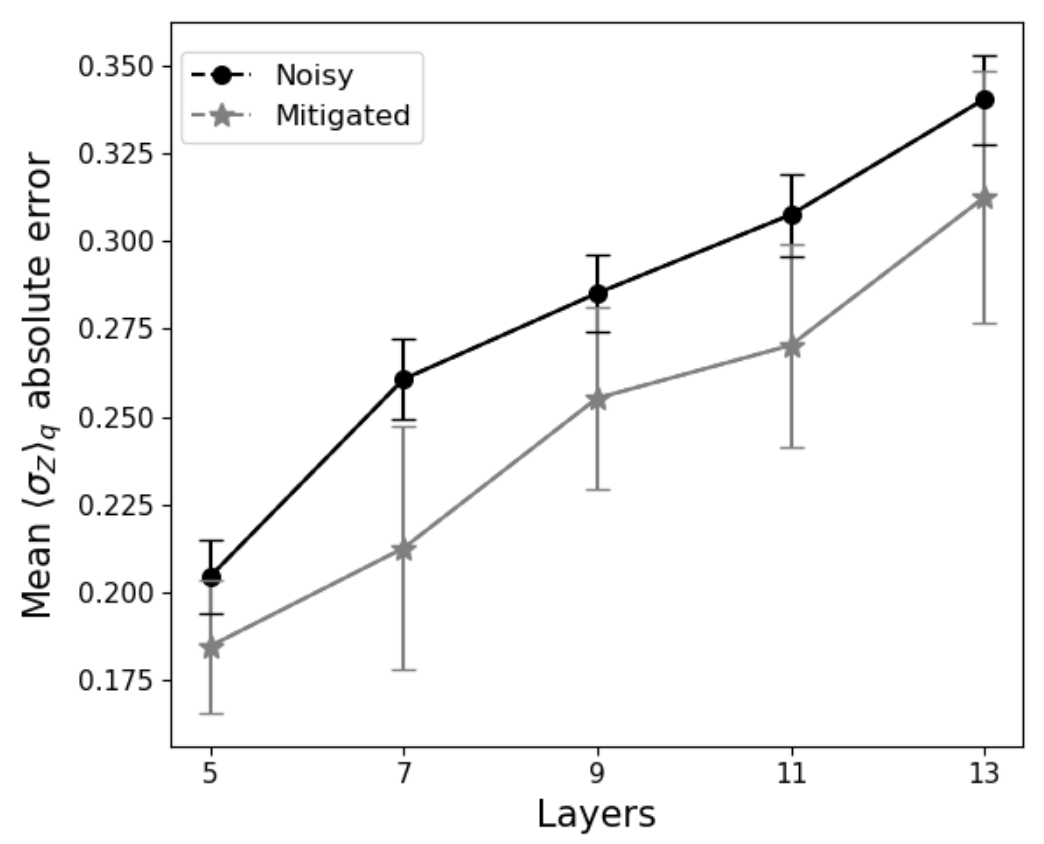}
\centering
\vspace*{-0.1cm} 
\caption{The mean absolute error of the expectation value of $\langle \sigma_{Z} \rangle$ from the local qubit measurements plotted against the number of circuit layers, with 95\% confidence interval error bars. Mitigating noise on the Aspen-9 Rigetti device with four qubits and varying numbers of gate layers. The circuits used for this were of the structure shown in Figure \ref{fig1}.}
\label{fig2}
\end{figure}

During sampling from the device, each time the target circuit is run the resulting output string is either accepted or rejected depending upon whether the acceptance threshold condition has been met in the trap circuit runs. The experiments were run across two four-qubit subsets on Aspen-9; these qubit subsets are labelled as 21, 22, 23, 24 and 32, 33, 34, 35 on the QPU specifications (see Figure \ref{fig4} (a)). Each qubit was measured in the computational basis, and the local expectation values output from the individual qubits were then averaged 

\begin{equation}
    \label{meanabsoluteerror1}
    < \sigma_{Z}>_{avg} =\sum_{q}\frac{1}{\abs{q}}< \sigma_{Z}>_q,
\end{equation}
where the sum is of the expectation values in the computational basis $< \sigma_{Z}>$ recorded for each qubit $q$. And so the mean absoluted error is 
\begin{equation}
    \label{meanabsoluteerror2}
    E_{abs} =|< \sigma_{Z}>^{ideal}_{avg}-< \sigma_{Z}>^{noisy}_{avg}|.
\end{equation}
As the readout is symmeterized by the quantum one-time pad we do not need to account for the asymmetric readout errors prevalent in superconducting hardware. Each experiment consists of 750 runs of the Accreditation protocol on each of the qubit subsets, resulting in a total of 1500 runs of the Accreditation protocol per data point. Each run of the protocol consists of 15 trap circuits and the target circuit, that is 16 circuit runs in total. The order of the trap and target circuits within one protocol run is randomly assigned. Fixing the number of qubits we ran the experiment with 5, 7, 9, 11 and 13 gate layers. For each of these respectively, a trap success cutoff of greater than 6, 6, 4, 4, and 4 was used. The trap success cutoff relates to the quality factor, defining the performance threshold $\varepsilon$ that has to be met for the target circuit output of a single run of Accreditation to be post-selected. For post-selection to occur, the number of trap circuit runs that do not record an error has to be greater than the designated trap success cutoff. These cutoff values were chosen heuristically depending on device performance for the particular circuit sizes. The corresponding numbers of post-selected output values were 439, 140, 238, 246 and 180. The results from this experiment can be observed in Figure \ref{fig2}, where the lower mean absolute errors for the post-selected sample group relative to the general sample group indicate a weaker overall noise channel in the former.

\subsection*{Quantum Circuit Born Machine Experiment}

\begin{figure*}[t]

\subfloat[]{%
  \includegraphics[scale=0.4]{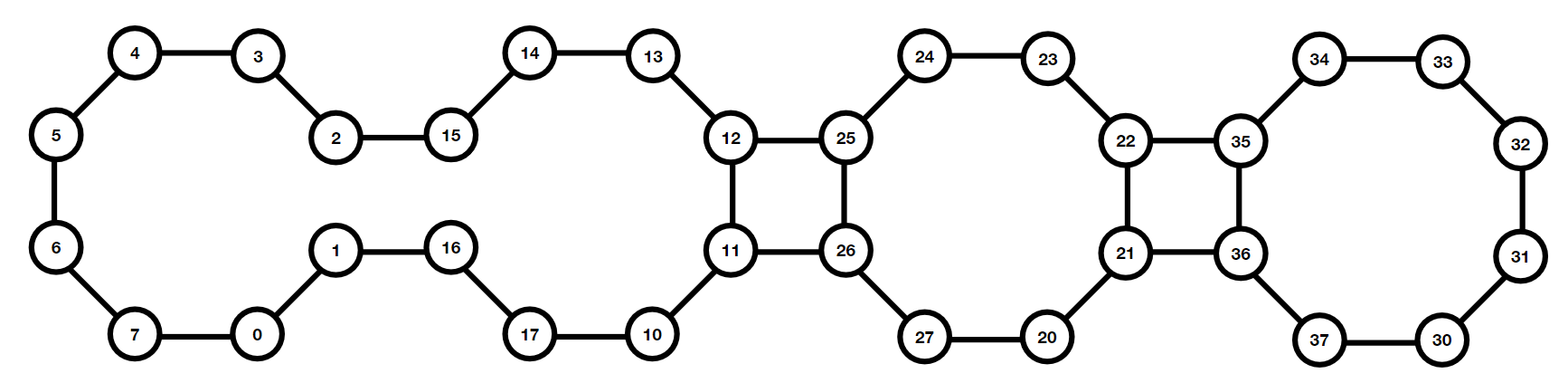}%
}

\subfloat[]{%
  \includegraphics[scale=0.65]{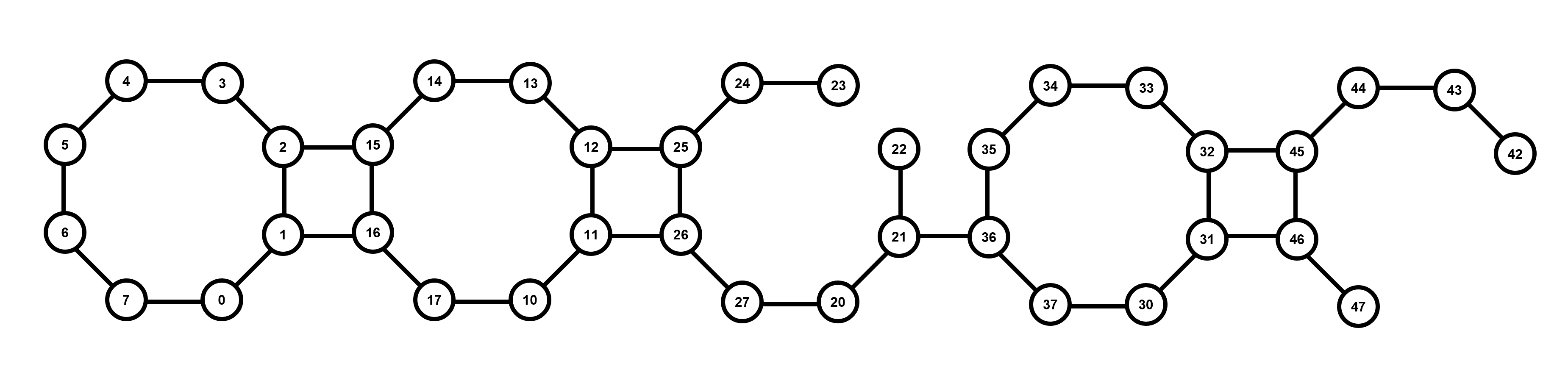}%
}
% \hspace*{-0.3cm}  
% \includegraphics[scale=0.5]{Aspen_11_topology.png}
% \centering
% \vspace*{-0.5cm} 
\caption{Quantum device topology diagrams displaying the on-chip qubit labels and connectivity, for: a) The Rigetti Aspen-9 32 qubit device, and b) The Rigetti Aspen-11 38 qubit device.}
\label{fig4}
\end{figure*}

A quantum circuit Born machine (QCBM) is a type of generative quantum machine learning model that can be used to produce new synthetic data according to some learned distribution. It has been previously considered as representing a possible candidate for quantum machine learning to achieve a performance advantage over comparable classical generative learning models \cite{coyle2020Bornmachine}. The QCBM has been demonstrated on currently available quantum hardware for some small practical use-cases, for instance in learning currency pair correlations in finance \cite{coyle2021finance}. Generative modelling of certain discrete known distributions has previously been utilised as a convenient benchmark for QCBM performance in the presence of errors \cite{hamilton2020Errormiti}. Here we apply the QCBM as a generative model to learn a discrete Poisson distribution, and use this to compare the closeness of the noisy to ideal distributions and also of the mitigated and ideal distributions.

To test the effect of our technique on the performance of a QCBM, a Poisson distribution was mapped onto the 16 available four-qubit computational basis states and this discrete distribution was then used as the target distribution during the training of a four-qubit, twenty-parameter QCMB on the Rigetti Aspen-11 QPU. 
%On this 38 qubit device, which has an octagonal topology displayed in \ref{fig4}, 
The four-qubit subsets 34, 35, 36, 37 and 42, 43, 44, 45 were used (see Figure \ref{fig4} (b)). 
%The device has a median T1 time of 29$\mu$s, median T2 time of 16$\mu$s, a median 1 qubit gate fidelity of 99.81\% and a median 2 qubit gate fidelity of 94.27\%.

\begin{figure*}[t]
%\hspace*{-1cm}  
\includegraphics[scale=0.36]{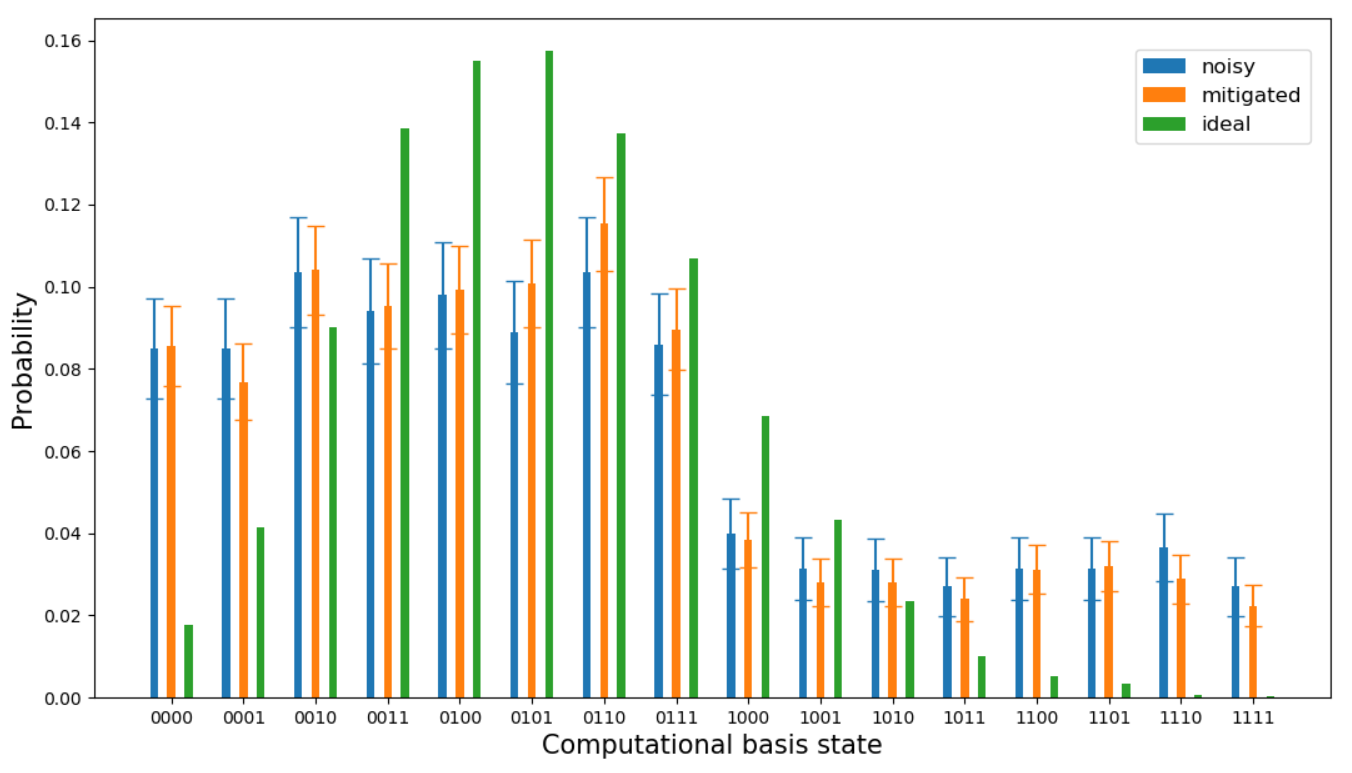}
\centering
%\vspace*{-0.8cm} 
\caption{The ideal, noisy and mitigated Poisson distributions shown mapped onto the 16 computational basis states, with 95\%  confidence interval error bars included for the noisy and mitigated distributions. The ideal Poisson distribution was used as the target distribution during training of the four-qubit, twenty-parameter quantum circuit Born machine. Applying the mitigation by verification technique during sampling from the trained circuit, the KL divergence for the unmitigated distribution was 0.624 and for the mitigated distribution was 0.531.}
\label{fig3}
\end{figure*}

The circuit structure for the QCBM was the same as that shown in Figure \ref{fig1} as used in the previous experiment, with alternating layers of local X rotation gates and $CZ$ gates. However rather than setting all of the X gate rotation parameters to $\pi$ like before, here the rotation parameters are trained by iterative optimization during a gradient descent algorithm which constitutes the circuit learning process. So that after training, sampling from the quantum circuit results in the output bitstrings occurring with frequencies dictated by a Poisson target distribution. The Poisson target distribution mapped onto the computational basis states is displayed Figure \ref{fig3}. 

The QCBM was trained for 100 iterations of the COBYLA optimization algorithm \cite{PowellCOBYLA}, with the Kullback-Leibler (KL) divergence \cite{KLdiv} used as the loss function during training. For each iteration of circuit training, 2000 samples were used to calculate the loss function which in this case was the KL divergence. The KL divergence is often used in the generative machine learning as a metric to represent the closeness of the noisy and ideal distributions \cite{Shim2018}. The KL divergence is defined as 
\begin{equation}
    \label{KLdivergence}
    D(p|q) =\sum_i p(x_i)\text{log}(\frac{p(x_i)}{q(x_i)})
\end{equation}
where $\{p(x)\}$ is the target distribution and $\{q(x)\}$ is the noisy distribution sampled from the QCBM. The performance metrics and loss functions used in training QCBMs vary widely in the literature, with choices being made optimally according to different circumstances and applications \cite{coyle2020Bornmachine}. However for the purposes of demonstrating our technique it was deemed sufficient to apply the commonly used KL divergence.

Our technique was used to mitigate the time-dependent errors (here corresponding to difference in noise levels between two subsets of qubits of a quantum device) in the noisy output distribution of the trained QCBM, with the distance from the ideal distribution calculated using the KL divergence. The trained QCBM was sampled 2000 times of which 1249 were post-selected using our mitigation technique. For each run of the protocol 16 traps were used and if greater than 4 traps ran correctly the output bitstring was post-selected. The KL divergence for the unmitigated distribution was 0.624 and for the mitigated distribution was 0.531, representing a reduction in the divergence of roughly 15\%. An ideal simulation of the QCBM in the absence of errors sampled 2000 times achieves a KL divergence of 0.191.

\section{\label{discussion}Discussion}
In summary, we have presented an error mitigation protocol for mitigating time-dependent errors which is based on the Accreditation protocol of \cite{ferracin2021experimentally} and post-selection. We have analyzed the sample complexity of our procedure, and shown that it provides a provable error mitigation for the important case of depolarizing noise. Finally, we ran our protocol on quantum hardware as an experimental proof of concept for the cases of a generic test circuit and a quantum circuit Born machine and obtained promising results.

A main advantage of our technique as compared to most other error mitigation techniques (see \cite{endo2021hybrid} for a review) is that it accounts for time-dependent noise behaviours. However, a weakness in our theoretical arguments is the assumption that this time-dependent behaviour of the noise can only be one of a $finite$ set of such behaviours. It would be interesting to look at ways in which we can overcome this assumption.

Furthermore, an interesting direction to pursue would be studying in detail how to integrate our error mitigation technique with other existing techniques dealing with noise which is not strongly time-dependent \cite{endo2021hybrid}. %Indeed, one might think of using our technique to filter-out experiments with very high errors, and keep only those  with low enough error, then use existing techniques  to lower errors  even further. 

Another important direction is adapting our protocol to work with other verification techniques, such as \cite{fitzsimons2017unconditionally} which can treat more general forms of noise including so-called \emph{malicious} noise coming from some adversary whose goal is to intentionally corrupt the computation.

We have run our protocol on Rigetti's Aspen-9 and Aspen-11 quantum processors, which are based on superconducting quantum technology. It would be interesting to test the performance of our protocol on other types of quantum technologies such as photonic  \cite{o2009photonic} or ion trap \cite{kielpinski2002architecture} technologies.

The main advantage of using the Accreditation protocol is its scalability \cite{ferracin2021experimentally,ferracin2019accrediting}. This makes our error mitigation protocol a promising candidate to be applied on quantum circuits of sufficiently high qubit number such as to allow a convincing demonstration of quantum advantage for specific problems. We aim to explore this direction in a future work.

%- \textbf{interpretation (convergence , sample complexity,...)}
%We first show that
%\begin{theorem}

%\end{theorem}
%-\textbf{talk abt time dependent assumptions (not needed for depolarizing noise, but needed in general that num. of behaviours is finite)}

%- twirling (add paulis in between) gives following form of noise, note that same prob dist of paulis across all targets and traps bcz identical 2 qubit gates, preps, meas, and single qubit gate noise is gate indep.

%- number of failed traps allows to upper bound error prob.

%-write equations, with confidence levels, and bla bla,...

%-in this work we introduce QEM with accreditation and post-selection

%-in accred. run multiple trap circuits with same noise behaviour as target, deduce from results of traps the quality of target (in terms of VD).

%-post select on the VD's which are below a certain value (quality factor)

%-this provably mitigates under realistic noise assumptions

%-compared to other techniques, we consider time dependent effects, as output can change from one run to another (note also exponential error suppression considers time dep effects), also allows for mitigating wider set of noise behaviour, which fall under assumptions of accred protocol

\begin{acknowledgments}
We thank Marco Paini, Mark Hodson, and Theodoros Kapourniotis for fruitful discussions. We would also like to thank Mark Hodson and Alex Hill for their assistance in obtaining comprehensive Aspen-9 and Aspen-11 device specifications. R.M and E.K are grateful for support from the grant Innovate UK Commercialising Quantum Technologies (application number: 44167). J.M is supported by grant funding from the University of Edinburgh, School of Informatics. We thank the anonymous Referees as well as the Editors  whose comments helped in improving this manuscript.
\end{acknowledgments}

\appendix
%\label{appendix}
\section{Appendixes\label{appendix}}
\subsection{ Proof of Theorem \ref{th1}}
Let $X_j \in \{p_{1}(+1)-p_{1}(-1),...,p_{\omega}(+1)-p_{\omega}(-1)\}$ for $j=1,...,m$ be independent identically distributed random variables, where for each $X_j$ every $p_{l}(+1)-p_{l}(-1)$ is chosen with uniform probability $\frac{1}{\omega}$. Recall that $p_{l}(+1)-p_{l}(-1)$ is  the expectation value $<O>_l$ (Equation (\ref{eqexpectationnoisebehaviourj})) corresponding to a noise behaviour with $TVD_l \leq \varepsilon$ (Equation (\ref{eqTVD2})), and $l=1,..,\omega$
ranges over all such behaviours. $s_{\omega}$ (Equation (\ref{eqsw})) can be rewritten as
$$s_{\omega}=\frac{1}{m}\sum_{j=1,...,m}X_j.$$
Note that $E(X_j)=<O>_{mit}=\frac{1}{\omega}\sum_{l=1,...,\omega}p_{l}(+1)-p_l(-1)$ for all $j=1,...,m$; $E(X_j)$ denotes the expectation value of $X_j$. Also, note that $VAR(X_j)=\sigma^2_{\omega}$ (Equation (\ref{eqvariance}) ) for all $j=1,..,m$, where $VAR(X_j)$ denotes the variance of $X_j$. We can now use Chebyshev's inequality \cite{saw1984chebyshev} straighforwardly to obtain 
\begin{equation}
    \label{eqchebyshev}
    Pr(|s_{\omega}-<O>_{mit}| \leq \varepsilon_1) \geq 1-\frac{\sigma^2_\omega}{m\varepsilon^2_1}.
\end{equation}
Setting $1-\frac{\sigma^2_\omega}{m\varepsilon^2_1} \geq \gamma$ we obtain
\begin{equation}
 \label{eqmapp}
 m \geq \frac{\sigma^2_\omega}{(1-\gamma)\varepsilon^2_1}.
\end{equation}
Recall that we run $K$ times the Accreditation protocol, and post-select over $m$ instances where $TVD \leq \varepsilon$, and that there are $N$ total noise behaviours each chosen uniformly and independently (assumption $A2$) with probability $\frac{1}{N}$; and out of these $\omega \leq N$ behaviours have $TVD_l \leq \varepsilon$. $K$ being large enough, and the distribution over noise behaviours being uniform, the law of large numbers estimates the number of runs $m$ where we obtained $TVD \leq \varepsilon$ as $$m \approx \frac{\omega}{N} K.$$ Using this and Equation (\ref{eqmapp}) we obtain that 
$$K \approx \geq \frac{N\sigma^2_\omega}{\omega(1-\gamma)\varepsilon^2_1}.$$ This completes the proof of Theorem \ref{th1}.
\subsection{Proof of Theorem \ref{th2}}
We will first consider the case where $s_\omega \geq 0$.
Our proof idea is based on a Chernoff bound argument \cite{chernoff1952measure}. Let $X$ be a random variable, $a \in \mathbb{R}$, and $s>0$. Our starting point is the following relation
\begin{equation}
    \label{eqchernoff}
Pr(X \geq a)=Pr(e^{sX} \geq e^{sa}) \leq \frac{E(e^{sX})}{e^{sa}}.
\end{equation}
Where the righmost part of Equation (\ref{eqchernoff}) comes from Markov's inequality.
Let $X = m<\tilde{O}>_{mit}=\sum_{j=1,..,m}\lambda_j$ (Equation (\ref{eqestimateomit})).
From the independence assumption $A2$ in section \ref{prelim}
\begin{equation}
\label{eqmoment1}
E(e^{s\sum_{j=1,..,m}\lambda_j})=\prod_{j=1,...,m}E(e^{s\lambda_j}).
\end{equation}
As seen previously, $\lambda_{j} \in \{-1,+1\}$ for $j=1,..,m$ is a random variable with $Pr(\lambda_j=+1)=p^{j}_{\pi(j)}(+1)$ and 
$Pr(\lambda_j=-1)=1-Pr(\lambda_j=+1)=p^{j}_{\pi(j)}(-1)$.

Let us now compute the moment generating function
\begin{equation}
\label{eqexpansion}
E(e^{s\lambda_j})=1+sE(\lambda_{j})+\frac{s^2E(\lambda^{2}_j)}{2!}+\frac{s^3E(\lambda^{3}_j)}{3!}+...
\end{equation}
It is easy to see that for $i \in \mathbb{N}$, $E(\lambda^{2i}_j)=1$
and $E(\lambda^{2i+1}_j)=p^{j}_{\pi(j)}(+1)-p^{j}_{\pi(j)}(-1)=2p^{j}_{\pi(j)}(+1)-1.$
Plugging these into Equation (\ref{eqexpansion}) we get
\begin{equation}
   \label{eqexpansion2}
   E(e^{s\lambda_j})=1+\frac{s^2}{2!}+\frac{s^4}{4!}+...+(2p^{j}_{\pi(j)}(+1)-1)(s+\frac{s^3}{3!}+\frac{s^5}{5!}+...)
\end{equation}
Note that $1+\frac{s^2}{2!}+\frac{s^4}{4!}+...=\frac{e^s+e^{-s}}{2}$ and $s+\frac{s^3}{3!}+\frac{s^5}{5!}+...=\frac{e^s-e^{-s}}{2}$. Plugging these into Equation (\ref{eqexpansion2}) and rearranging we obtain
$$E(e^{s\lambda_j})=\frac{1+p^{j}_{\pi(j)}(+1)(e^{2s}-1)}{e^s}.$$
Using the fact that $1+\alpha \leq e^{\alpha}$ for all $\alpha \in \mathbb{R}$, and setting $\alpha=p^{j}_{\pi(j)}(+1)(e^{2s}-1)$
we obtain
\begin{equation}
    \label{eqmomentfinal}
    E(e^{s\lambda_j})\leq e^{p^{j}_{\pi(j)}(+1)(e^{2s}-1)-s}.
\end{equation}
Plugging this into Equation (\ref{eqmoment1}) we obtain

\begin{equation}
    \label{eqproductmomentfinal}
    E(e^{s\sum_{j=1,..,m}\lambda_j}) \leq e^{(e^{2s}-1)\sum_{j=1,,,m}p^{j}_{\pi(j)}(+1)-ms}.
\end{equation}
plugging this into Equation (\ref{eqchernoff}), and then flipping this Equation (i.e considering $Pr(X \leq a)$)  with $X=\sum_{j=1,...,m}\lambda_j=m<\tilde{O}>_{mit}$ we get
\begin{equation}
    \label{eqchernoff2}
    Pr(m<\tilde{O}>_{mit} \leq a) \geq 1-e^{(e^{2s}-1)\sum_{j=1,,,m}p^{j}_{\pi(j)}(+1)-ms-sa}.
\end{equation}
Choosing $a=(1+\delta)ms_{\omega}$ and plugging this into Equation (\ref{eqchernoff2}) we obtain
\begin{equation}
   \label{eqchernoff3}
   Pr(<\tilde{O}>_{mit} \leq (1+\delta)s_{\omega}) \geq 1-e^{f(s)},
\end{equation}
where $f(s)=(e^{2s}-1)u_1-(m+(1+\delta)u_2)s,$ we have relabelled (for convenience) $u_1=\sum_{j=1,...,m}p^{j}_{\pi(j)}(+1)$
and $u_2=ms_{\omega}=2u_1-m$. 
In order to make our probability for the bound as large as possible, we will minimize $f(s)$ over $s$. Computing the derivative $\frac{df(s)}{ds}$ and setting it equal to 0, we get the following value for $s$ which maximizes $1-e^{f(s)}$
\begin{equation}
  \label{eqmins}
  s_{min}=\frac{1}{2}ln(\frac{m+(1+\delta) u_2}{2u_1}).
\end{equation}
Note that we require $s\geq 0$, which means $\frac{m+(1+\delta) u_2}{2u_1} \geq 1$ which implies (after replacing $u_2=2u_1-m$ and solving) that $u_1 \geq \frac{m}{2}$.
This means that $2u_1-m \geq 0$ and therefore $s_{\omega} \geq 0$ (as we required in the beginning of this proof case). 
Replacing $s_{min}$ in Equation (\ref{eqchernoff3}) and rearranging we obtain
\begin{equation}
    \label{eqchernoff4}
    Pr(<\tilde{O}>_{mit} \leq (1+\delta)s_{\omega}) \geq 1-e^{u_1f(y)}
\end{equation}
with $f(y)=y(1-ln(y))-1 \leq 0$  and $y=1+\delta-\frac{\delta m}{2u_1} >0.$
Note that, because $\frac{m}{2} \leq u_1 \leq m$ then $0<\delta < y \leq 1+\frac{\delta}{2}$.
Now, for good accuracy, we require $\delta$ to be small, so $f(y) \approx (1+\delta-\frac{\delta m}{2u_1})(1-(\delta-\frac{\delta m}{2u_1}))-1 \approx -\delta^2(1-\beta)^2$, where $\beta:=\beta(m)=\frac{m}{2u_1}$. Note that $1/2\leq \beta<1$, and also that $s_\omega=\frac{1}{\beta}-1$, and from Theorem \ref{th1} we know that $s_\omega$ converges to a fixed value (i.e $<O>_{mit}$) then also $\beta$ converges to a fixed value.

Now, $u_1f(y)=\frac{u_1}{m}mf(y) \approx -\frac{1}{2\beta}(1-\beta)^2\delta^2m$. Setting $\frac{1}{2\beta}(1-\beta)^2\delta^2m > l$ and using the bound on $m$ found in Equation (\ref{eqmapp}), we can choose
$\delta \approx \frac{\sqrt{2\beta} }{1-\beta}\sqrt{\frac{l\varepsilon^{2}_1(1-\gamma)}{\sigma^2_{\omega}}}.$ This proves Equation (\ref{eqth21}). 

Note that for the case where  $\beta=1$ ($s_{\omega}=0$), $u_1f(y)=0$ since $y=1$. Also, $(1 \pm \delta)s_w$ can be made continuous by extension at $\beta=1$ since 
\begin{multline*}
lim_{\beta \to 1}(1 \pm \delta)s_{\omega}=lim_{\beta \to 1}(1 \pm \frac{\sqrt{2\beta} }{1-\beta}\sqrt{\frac{l\varepsilon^{2}_1(1-\gamma)}{\sigma^2_{\omega}}})(\frac{1}{\beta}-1)=\\ lim_{\beta \to 1} \frac{1-\beta}{\beta} \pm \frac{\sqrt{2\beta}}{\beta}\frac{1-\beta}{1-\beta}\sqrt{\frac{l\varepsilon^{2}_1(1-\gamma)}{\sigma^2_{\omega}}}=\pm \sqrt{2\frac{l\varepsilon^{2}_1(1-\gamma)}{\sigma^2_{\omega}}}.
\end{multline*}
This leads to the (trivial) bounds \begin{equation*}
   %\label{eqchernoff3}
   Pr\big( <\tilde{O}>_{mit} \leq \sqrt{2\frac{l\varepsilon^{2}_1(1-\gamma)}{\sigma^2_{\omega}}} \big)   \geq 0,
\end{equation*}
and
\begin{equation*}
   %\label{eqchernoff3}
   Pr \big( <\tilde{O}>_{mit} \geq -\sqrt{2\frac{l\varepsilon^{2}_1(1-\gamma)}{\sigma^2_{\omega}}} \big)   \geq 0.
\end{equation*}

The proof of Equations (\ref{eqth22}), (\ref{eqth23}) and (\ref{eqth24}) follows a similar path, so we will not repeat the calculations, but will remark on small differences in proving these as compared to proving Equation (\ref{eqth21}). In Equation (\ref{eqth22}) the starting point is the inequality $Pr(X \leq a) \leq\frac{ E(e^{-sX})}{e^{-sa}}$ also obtained from a Markov inequality, and we choose $a=(1-\delta)ms_\omega$. In proving Equations (\ref{eqth23}) and (\ref{eqth24}) we have, $s<0$, $u_1 < \dfrac{m}{2}$ (i.e $s_\omega <0$), and $1<\beta \leq \frac{1}{2\lambda}$. For Equation (\ref{eqth23}) the starting point is $Pr(X \geq a) \leq \frac{E(e^{-sX})}{e^{-sa}}$ and $a=(1-\delta')ms_\omega$. For Equation (\ref{eqth24}) the starting point is $Pr(X \leq a) \leq \frac{E(e^{sX})}{e^{sa}}$, and $a=(1+\delta')ms_\omega$.
Also in Equation (\ref{eqth24}) in order for $f(y)$
with $y=1+\delta'-\frac{\delta'm}{2u_1}$ to be well-defined, we must have $y>0$ which imposes the condition (note that $1<\dfrac{m}{2u_1}=\beta \leq \frac{1}{2\lambda}$)
$$0<\delta'< \frac{2\lambda}{1-2\lambda}.$$ This condition is what leads to the constraint on $l'$ in Theorem \ref{th2}. This completes the proof of Theorem \ref{th2}.

\section{Aspen-9 and Aspen-11 QPU Specification Information}
\label{appB}
\begin{figure*}[t]

\subfloat[]{%
  \includegraphics[scale=0.29]{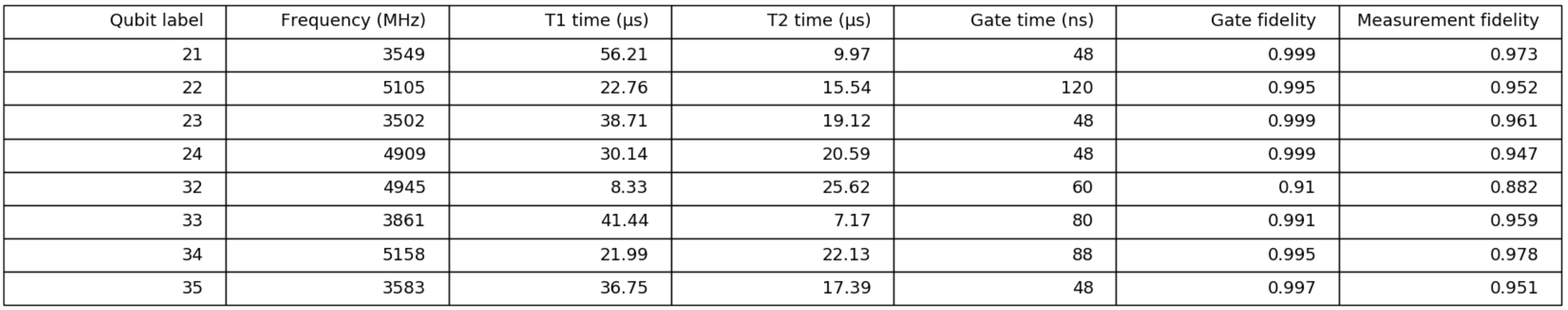}%
}

\subfloat[]{%
  \includegraphics[scale=0.29]{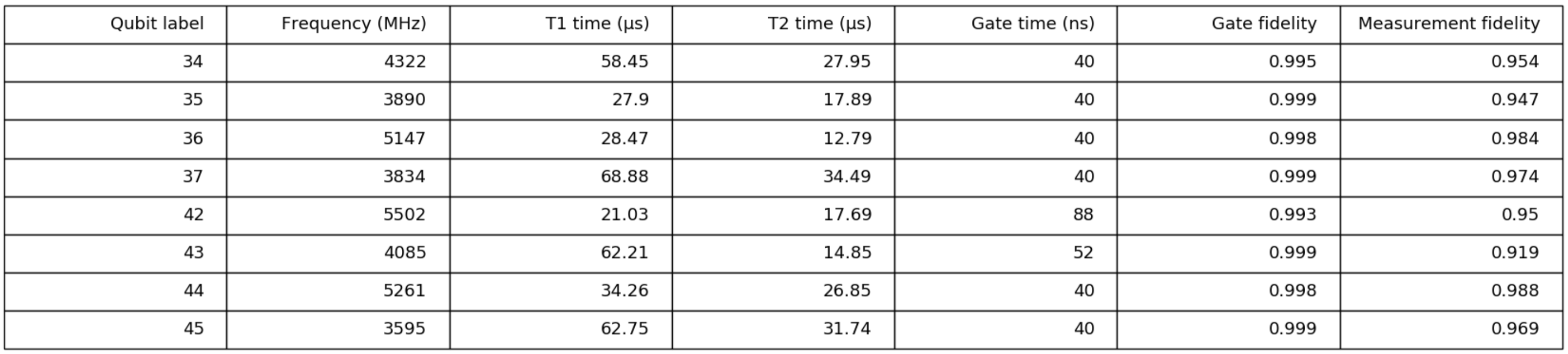}%
}
\caption{Quantum device specifications, for: a) The Rigetti Aspen-9 32 qubit device, and b) The Rigetti Aspen-11 38 qubit device.}
\label{fig5}
\end{figure*}

Information on the device specifications for the Rigetti Aspen-9 and Aspen-11 QPUs is included in Figure \ref{fig5} a) and b) respectively. The Aspen-9 experiments were performed on 25th September 2021, and the Aspen-11 experiments were performed on 5th February 2022. For each of the qubits used in our experiments, and for each device, we report information on qubit frequency, T1 and T2 time, gate time, gate fidelity and measurement fidelity.

\bibliography{apssamp.bib}% Produces the bibliography via BibTeX.

\end{document}